\documentclass[sn-nature]{sn-jnl}

\usepackage{graphicx}%
\usepackage{multirow}%
\usepackage{amsmath,amssymb,amsfonts}%
\usepackage{amsthm}%
\usepackage{mathrsfs}%
\usepackage[title]{appendix}%
\usepackage{xcolor}%
\usepackage{textcomp}%
\usepackage{manyfoot}%
\usepackage{booktabs}%
\usepackage{algorithm}%
\usepackage{algorithmicx}%
\usepackage{algpseudocode}%
\usepackage{listings}%

\theoremstyle{thmstyleone}%
\theoremstyle{thmstyletwo}%

\theoremstyle{thmstylethree}%

\newcommand{\CdAs}{$\mathrm{Cd_{3}As_{2}}$}
\newcommand{\InMnAs}{$\mathrm{In_{1-x}Mn_{x}As}$}
\newcommand{\GaMnSb}{$\mathrm{Ga_{1-x}Mn_{x}Sb}$}
\newcommand{\GaMnAs}{$\mathrm{Ga_{1-x}Mn_{x}As}$}
\newcommand{\Tc}{$T_{\mathrm{C}}$}
\newcommand{\Vg}{$V_{\mathrm{g}}$}
\newcommand{\rhoxx}{$\rho_{\mathrm{xx}}$}
\newcommand{\rhoxy}{$\rho_{\mathrm{xy}}$}

\begin{document}

\title[Efficiently gate-tunable ferromagnetism in ferromagnetic semiconductor-Dirac semimetal p-n heterojunctions]{Efficiently gate-tunable ferromagnetism in ferromagnetic semiconductor-Dirac semimetal p-n heterojunctions}


\author[1]{\fnm{Emma} \sur{Steinebronn}}

\author[1]{\fnm{Saurav} \sur{Islam}}

\author[1]{\fnm{Abhinava } \sur{Chatterjee}}

\author[2]{\fnm{Bimal } \sur{Neupane}}

\author[3] {\fnm{Alex } \sur{Grutter}}

\author[3] {\fnm{Christopher} \sur{Jensen}}

\author[3] {\fnm{Julie A. } \sur{Borchers}}

\author[4] {\fnm{Timothy} \sur{Charlton}}

\author[1] {\fnm{Wilson J. } \sur{Y\'{a}nez-Parre\~{n}o}}

\author[5] {\fnm{Juan} \sur{Chamorro}}

\author[5] {\fnm{Tanya} \sur{Berry}}

\author[6] {\fnm{Supriya } \sur{Ghosh}}

\author[1] {\fnm{K. A.} \sur{Nivedith}}

\author[6] {\fnm{K. Andre} \sur{Mkhoyan}}

\author[5] {\fnm{Tyrel } \sur{McQueen}}

\author[2] {\fnm{Yuanxi } \sur{Wang}}

\author[1] {\fnm{Chaoxing} \sur{Liu}}

\author[1,7,8] {\fnm{Nitin} \sur{Samarth}}

\affil[1]{\orgdiv{Department of Physics}, \orgname{Pennsylvania State University}, \orgaddress{\city{University Park}, \postcode{16802}, \state{PA}, \country{USA}}}

\affil[2]{\orgdiv{Department of Physics}, \orgname{University of North Texas}, \orgaddress{\city{Denton}, \postcode{76203}, \state{TX}, \country{USA}}}

\affil[3]{\orgname{National Institution of Standards and Technology}, \city{Gaithersburg}, \postcode{20899}, \state{MD}, \country{USA}}

\affil[4]{\orgname{Oak Ridge National Lab}, \orgaddress{\city{Oak Ridge}, \postcode{37830}, \state{TN}, \country{USA}}}

\affil[5]{\orgdiv{Department of Chemistry}, \orgname{Johns Hopkins University}, \orgaddress{\city{Baltimore}, \postcode{21218}, \state{MD}, \country{USA}}}

\affil[6]{\orgdiv{Department of Chemical Engineering and Materials Science}, \orgname{University of Minnesota}, \orgaddress{\city{Minneapolis}, \postcode{55455}, \state{MN}, \country{USA}}}

\affil[7]{\orgdiv{Department of Materials Science and Engineering}, \orgname{Pennsylvania State University}, \orgaddress{\city{University Park}, \postcode{16802}, \state{PA}, \country{USA}}}

\affil[8]{\orgdiv{Materials Research Institute}, \orgname{Pennsylvania State University}, \orgaddress{\city{University Park}, \postcode{16802}, \state{PA}, \country{USA}}}


\abstract{We use molecular beam epitaxy to develop a gate tunable p-n heterojunction that interfaces a canonical Dirac semimetal, Cd$_3$As$_2$, and a ferromagnetic semiconductor, In$_{1-x}$Mn$_x$As, with perpendicular magnetic anisotropy. Measurements of the anomalous Hall effect in top-gated Cd$_3$As$_2$/In$_{1-x}$Mn$_x$As devices show that the ferromagnetic Curie temperature ($T_\mathrm{C}$) can be efficiently tuned using a modest gate voltage of $\sim 10$ V, corresponding to a sensitivity to electric field ($E$) of $\Delta T_{\mathrm{C}}/\Delta E \sim 10$ K/MV/cm). The voltage tuning of $T_\mathrm{C}$ saturates near the charge neutrality point of Cd$_3$As$_2$~and vanishes at positive gate voltage in appropriately designed heterostructures. This non-monotonic behavior cannot be explained solely by hole-mediated ferromagnetism in the In$_{1-x}$Mn$_x$As alone, suggesting an interaction between the Dirac semimetal and the ferromagnetic semiconductor. Our results identify Cd$_3$As$_2$/In$_{1-x}$Mn$_x$As heterojunctions as a potentially attractive platform for studying emergent phenomena arising from the interplay between broken symmetry, topology, and magnetism in a topological semimetal. }




\maketitle

\section*{Introduction}\label{sec1}

Topological materials have attracted considerable interest in the past decade, exhibiting symmetry-protected electronic states due to the complex interplay of lattice structure, symmetry, and spin-orbit coupling~\cite{hasan2010colloquium,ando2013topological}. Dirac semimetals (DSMs) form an important part of the topological material landscape~\cite{burkov2016topological,yan2017topological,armitage2018weyl}: their hallmark behavior arises from spin-degenerate Dirac fermions protected by both time-reversal symmetry (TRS) and inversion symmetry, yielding the promise of phenomena such as topological superconductivity\cite{Kobayashi_PhysRevLett.115.187001}, the chiral anomaly \cite{Parameswaran_2014}, and the axial magnetoelectric effect\cite{PhysRevLett.126.247202}. When one of these symmetries is broken, Kramer's degeneracy is lifted, separating these four-fold degenerate Dirac fermions into two-fold degenerate Weyl fermions of opposite chirality and resulting in a magnetic Weyl semimetal (WSM). Although several materials have been identified as intrinsic WSMs \cite{WSM_1, WSM_2, WSM_3, WSM_4}, we seek to design a material system that could be tuned between the DSM and WSM topological phases using an external parameter such as temperature, magnetic field, or electric field. Such a heterostructure could serve as a model platform for systematic studies of topological phase transitions. It could also provide an attractive route toward a platform to fine-tune a monopole superconductor state in hybrid superconductor/topological semimetal Josephson junction geometries \cite{bobrow2020monopole,bobrow2022monopole}. Previous studies have attempted to break TRS in a DSM via magnetic doping of thin films of the canonical DSM, \CdAs, using molecular beam epitaxy (MBE) \cite{XiaoMnDoping}. However, careful transmission electron microscopy (TEM) revealed that the Mn dopants floated to the top, rather than being homogeneously interspersed throughout the \CdAs~layer. (A recent report shows a growth protocol that can avoid such segregation \cite{RiceMnDoping}.)
Here, we develop and characterize a hybrid DSM/ferromagnetic (FM) semiconductor heterojunction aimed at realizing an electrically-tunable transition from a DSM to a WSM by inducing FM order in Cd$_3$As$_2$ via the magnetic proximity effect. 
We recently attempted such an approach in [111] oriented \CdAs/\GaMnSb~heterostructures but found no compelling evidence for induced ferromagnetism in the DSM down to temperatures as low as $T = 4$~K~\cite{Mitra2023}. This lack of success was mainly attributed to the in-plane magnetic anisotropy of \GaMnSb, as well as its inhomogeneous ferromagnetism.

Here, we use an alternative approach by interfacing \CdAs~with another well-studied FM semiconductor, \InMnAs, with a robust perpendicular magnetic anisotropy (PMA) when grown on GaAs (001) (see schematic in Fig. \ref{Figure 1} (a)).   \cite{ohno2000electric,Dietl_RevModPhys.86.187,Sinova_RevModPhys.78.809}. Since \InMnAs~is a p-type semiconductor and MBE-grown \CdAs~films have chemical potential in the conduction band above the Dirac point \cite{schumann2016molecular,schumann2018observation,Xiao_PhysRevB.106.L201101,kealhofer2020topological}, this combination of materials forms a p-n heterojunction. Optimal epitaxial growth of \CdAs~films with reasonable structural quality and high mobility on (001) GaAs or GaSb substrates typically produces carrier densities of $\sim 10^{12} {\textrm{cm}}^{-2}$ in as-grown films~\cite{kealhofer2020topological}. In the \CdAs/\InMnAs~heterojunctions studied here, we are interested in modulating the hole density ($\sim 10^{14} {\textrm{cm}}^{-2}$) in the \InMnAs~so as to tune the hole-mediated ferromagnetism. Our \CdAs~films are grown under sub-optimal conditions, creating defects that lead to an excess extrinsic carrier density that closely matches the hole density in the vicinal FM semiconductor. The disorder consists of diffusion regions at the interfaces and grain boundaries, as seen in the high resolution transmission electron microscopy (HRTEM) cross-sectional image of a sample shown in Fig. \ref{Figure 1} (b) (additional TEM data are available in the Supplementary Information). Note that at the relatively small Mn concentration used here, \CdAs/\InMnAs~has a lattice mismatch of about 4.6\%, similar to the case of \CdAs/GaSb~ \cite{kealhofer2020topological}.  Thus, the disorder and excess carrier density is not a result of the lattice mismatch.

Like earlier studies of hybrid FM semiconductor/topological insulator~\cite{lee2018engineering} or FM semiconductor/semimetal heterostructures~\cite{Mitra2023}, we select a Mn composition ($x \sim 2 \%$) that maximizes the resistivity of the III-Mn-V FM semiconductor while still allowing FM ordering. This minimizes the effects of a shunting path through the FM semiconductor when an electric current passes through a bilayer \InMnAs/\CdAs~device. We then measure the anomalous Hall effect (AHE) in top-gated \CdAs/\InMnAs~heterojunctions to demonstrate gate voltage (\Vg)-tunable ferromagnetism with a large change in the FM Curie temperature (\Tc). In an appropriately designed heterojunction, using a modest variation in gate voltage ($-6$ V $\leq$ \Vg $\leq$ $+6$ V), corresponding to a maximum electric field of about 1 MV/cm, we control \Tc~from completely quenched FM order to a maximum value of about 12 K. The results we present here are qualitatively different from previous studies of electrically-tunable ferromagnetism in III-Mn-V FM semiconductors \InMnAs~\cite{ohno2000electric} and \GaMnAs~\cite{Chiba_APL_2006} where modest monotonic changes in \Tc~(a few K) were observed by modulating the hole density in gated FM semiconductor devices. The hybrid FM semiconductor/DSM p-n heterojunctions studied here show a non-monotonic variation of \Tc~with \Vg~and a correlation with the charge neutrality point (CNP). This suggests an important role played by the exchange interaction between the Dirac electrons in the DSM and the magnetic moments in the FM semiconductor. As shown elsewhere \cite{Islam_arxiv}, this interaction may also be responsible for the creation of interfacial chiral spin texture in \CdAs/\InMnAs~heterojunctions at lower temperatures than discussed here.

\section*{Results}\label{sec2}

\subsection*{Band structure calculations}
\begin{figure}
    \includegraphics[width=12cm]{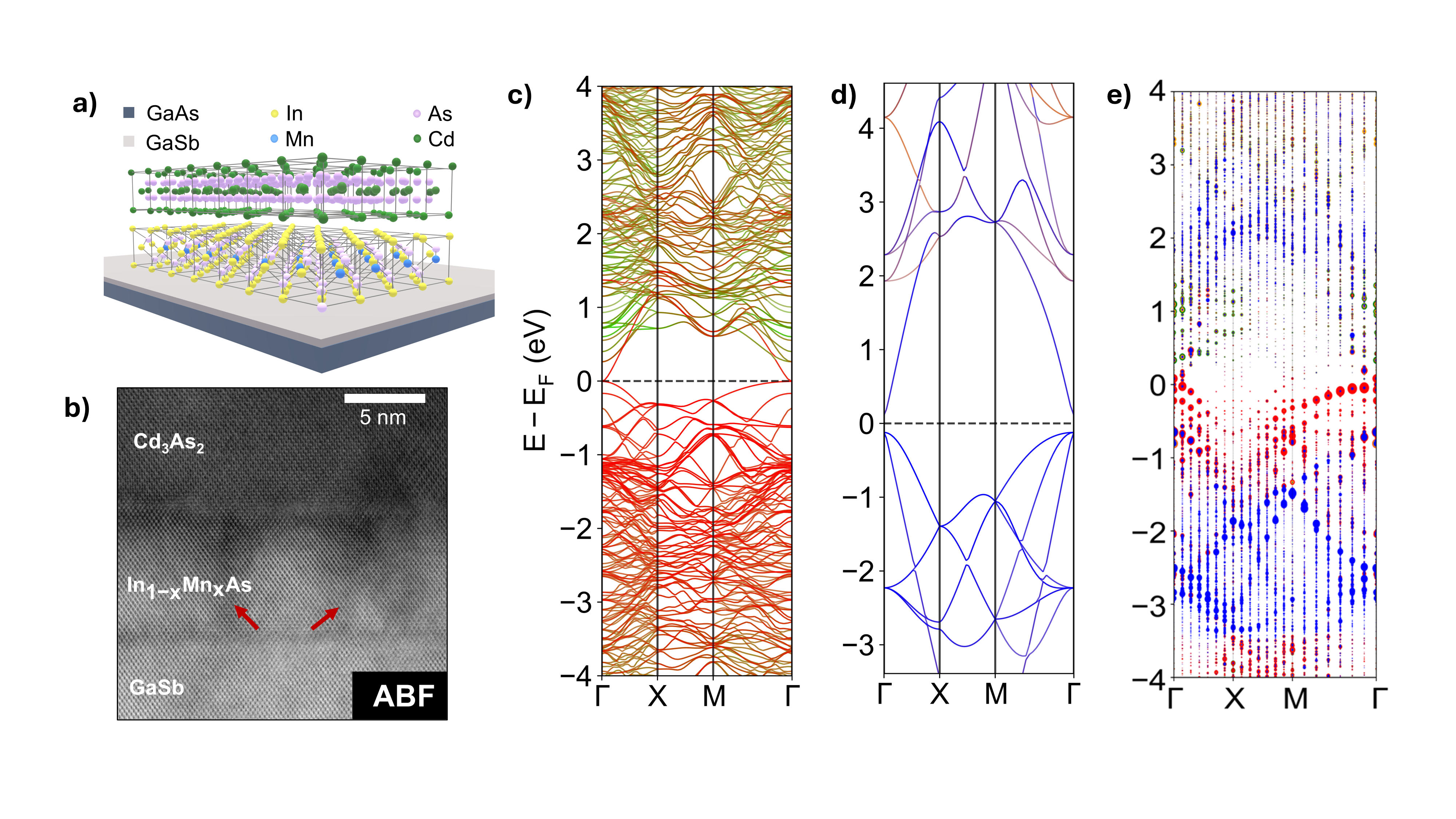}
    \caption{\textbf{Proposed Heterostructure:}  (a) Schematic of heterostructure grown. (b) Cross-sectional annular bright field scanning transmission electron microscopy (ABF-STEM) image of a \CdAs/\InMnAs~heterostructure. The vertical dark regions marked by red arrows in the \InMnAs~layer are indicative of grain boundaries. (c)-(d) Orbital-projected band structures for (c) a conventional unit cell of \CdAs~ and (d) a conventional unit cell of InAs. (e) Unfolded band structure of the \CdAs/InAs heterostructure, projected onto the $k$-point path of the conventional unit cell of InAs. 
    }
      \label{Figure 1}
 \end{figure}

Before describing the experimental results, we discuss the band structure of \CdAs/\InMnAs. To locate the \CdAs~Dirac point (DP) relative to the InAs bandgap and determine potential charge transfer across the \CdAs/InAs interface, we perform first-principles calculations using a heterostructure model connecting a $2\times2\times2.5$ supercell of a conventional InAs unit cell with one conventional unit cell of \CdAs~along the $[001]$ direction. For the Mn-doping levels in the III-Mn-V FM semiconductor studied here (a few percent), the changes in band structures are expected to be insignificant,  with the principal anticipated impact being a reduction of the Fermi level due to the p-type doping. This leads us to carry out theoretical calculations of just the \CdAs/InAs band alignment. (Note that explicit inclusion of Mn doping would also be computationally far more intensive.) We first estimate band alignment by matching atomic core levels in the heterostructure model with those in pristine bulk \CdAs~ and InAs~\cite{Huai}. Specifically, we match the 1\textit{s} orbital core eigenvalues of As in the \CdAs~ (InAs) region of the heterostructure with those in pristine \CdAs~(InAs). The resulting aligned  band structures of the three systems are shown in Fig. \ref{Figure 1} (c-e), where the Fermi energies are all set to zero. We observe that \CdAs~DP is positioned $0.73$~eV above the InAs valence band maximum and $0.49$~eV above the InAs conduction band minimum, suggesting electron transfer from \CdAs~to InAs. To ensure convergence with respect to the layer thickness of the modeled slab, we tested slab thicknesses from $25.92$~Å to $45.31$~Å, where the band offset was converged within $0.03$~eV at a thickness of $32.87$~Å. 

Taking a second perspective at estimating band alignment, we next directly inspect the band structure of the heterostructure model by unfolding~\cite{PhysRevLett.104.216401} bands of the \CdAs/InAs heterostructure onto the Brillouin zone of the InAs conventional unit cell, along with orbital projection. In the unfolded bands in Fig. \ref{Figure 1} (e), the red valence bands originate from As atoms in \CdAs, and the blue bands originate from As atoms in bulk InAs. The same projection color scheme is also applied to the pristine \CdAs~and InAs band structures in Fig. \ref{Figure 1} (c,d). The fact that the InAs-dominated bands and \CdAs-dominated band in Fig. \ref{Figure 1} (e) are already aligned with those in Fig. \ref{Figure 1} (c,d) indicates that our previous band alignment procedure was an accurate approximation, and that interfacial covalent bonds do not significantly distort the band structure of each component. The conclusion is similar to our first band alignment attempt: the \CdAs-dominated valence band top (where the DP is expected) are positioned $0.73$~eV above the InAs valence band maximum in Fig. \ref{Figure 1} (e). We note that the DP cannot be exactly located in the heterostructure model along $\Gamma-z$; this is because recovering the DP by approaching the bulk limit requires thicknesses that are not practical with existing computational resources. For example, even with a doubled calculation cell, a moderate gap would still be expected for \CdAs~due to quantum well effects. Hence, we choose to follow the \CdAs-dominated valence band maximum instead of the DP.

\begin{figure*}
    \includegraphics[width=12cm]{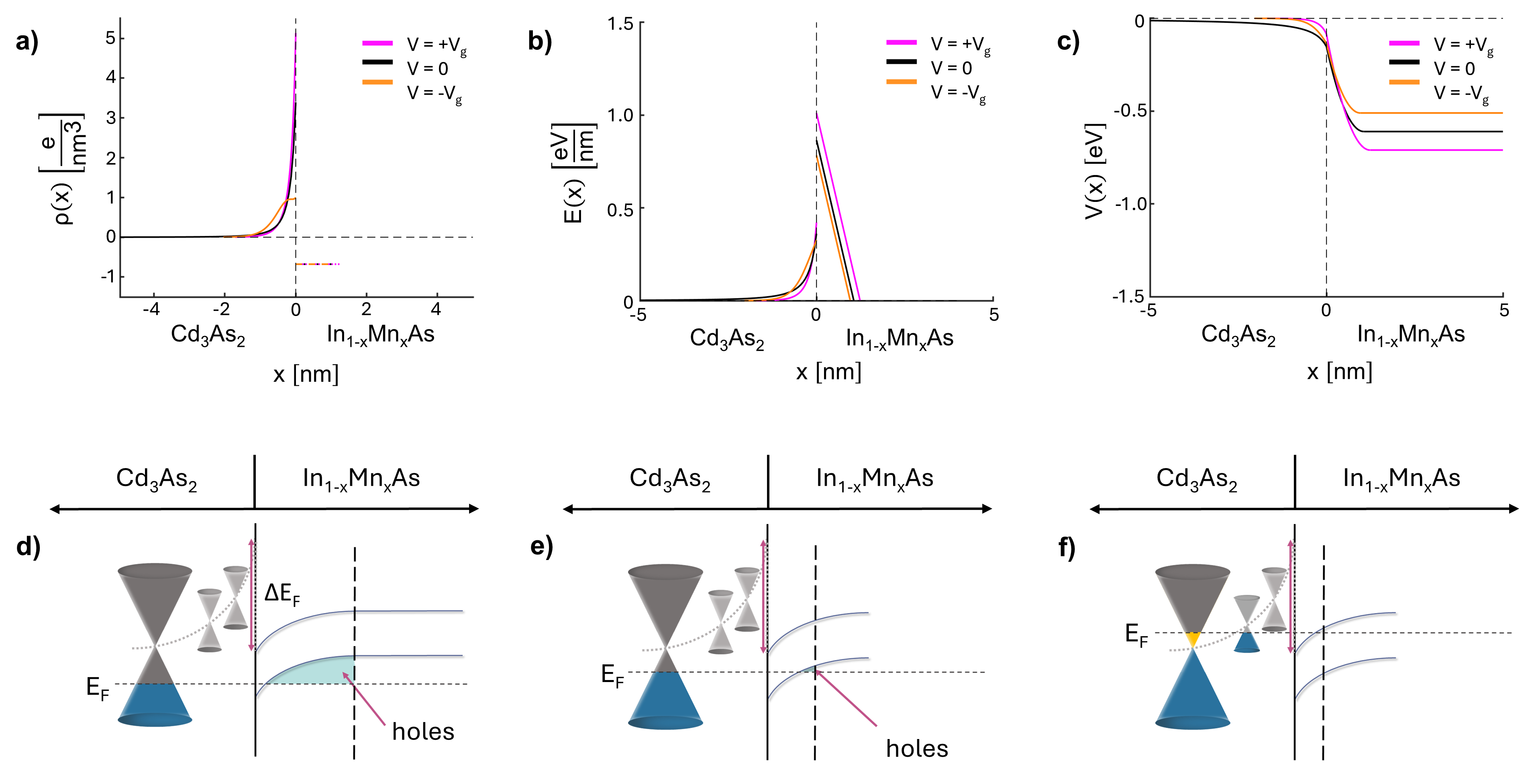}
    \caption{\textbf{Band Diagrams :} (a-c) Self-consistent solutions for charge density, electric field, and voltage plotted as a function of sample depth. (d-f) Depletion region  band alignment cartoons showing the effect of varying \InMnAs~thickness. Note as thickness of \InMnAs~decreases, the presence of holes is reduced, allowing the presence of holes to be easily controlled by tuning the chemical potential via a gate voltage.}
    \label{2}
 \end{figure*}

We use these data to develop a picture of band-bending across the heterostructure given the p-n (\InMnAs~ and \CdAs, respectively) nature of the material system. The qualitative description of the band bending is presented in Fig. \ref{2}. To better understand the DSM-FM semiconductor heterostructure, we present a simple p-n junction model that captures the essential features of the band bending across the heterostructure. Since In$_{1-x}$Mn$_x$As is a III-V semiconductor, we assume a constant charge density on the In$_{1-x}$Mn$_x$As side (p-side). As the \CdAs~side (n-side) is a DSM, we solve the full Poisson equation:  

\begin{equation}
    \phi_n''(x) \equiv \frac{d^2\phi_n(x)}{dx^2} = A \left( V_g^3 - \left( V_g - \phi_n(x)\right)^3 \right), \quad A = \frac{1}{3 \pi^2 \hbar^3 v_f^3 \epsilon_n},
\end{equation}
where $\phi_n$ is the electric potential on the Cd$_3$As$_2$ side, $v_f$ and $\epsilon_n$ are the Fermi velocity and dielectric constant of Cd$_3$As$_2$, respectively, and $V_g$ is the gate voltage (see Supplementary Section II for clarification). We take the potential to be zero far away on the Cd$_3$As$_2$ side and to be the built-in potential $-\phi_{bi}$, far away on In$_{1-x}$Mn$_x$As side, such that across the heterostructure, the potential change is $-\phi_{bi}$, which is the energy needed to match the Fermi levels. Naturally, $\phi_{bi}$ is given by the difference of the intrinsic Fermi levels between Cd$_3$As$_2$ and In$_{1-x}$Mn$_x$As. If the Fermi level on the Cd$_3$As$_2$ side is at the Dirac point and the Fermi level on the In$_{1-x}$Mn$_x$As side is mid-gap, then $\phi_{bi} = 0.61$ eV.  We also require that the electric field vanishes at the edges of the depletion region. The charge density, electric field, and electric potential in the depletion region is shown in Fig. \ref{2} (a-c). We observe a non-monotonic behavior in terms of the gate voltage $V_g$. 

While the electronic behavior is gate-tunable, we also note that the hypothesized proximity-induced magnetism in \CdAs~should be proportional to the magnetization in the \InMnAs~layer. To develop a platform in which one may switch between DSM and WSM states, one must be able to quench the FM order in the \InMnAs~layer. This may be accomplished by inducing sufficient electron-hole recombination using an applied voltage. By reducing the thickness of the \InMnAs~we hypothesize the development of such a platform as visualized in Fig. \ref{2} (d-f).

 \subsection*{MBE growth and characterization}
We grow thin films of \InMnAs~and \CdAs, and \CdAs/\InMnAs~heterostructures using MBE on epi-ready GaAs (001) substrates (see Methods and Supplementary Materials Section III for details). The structure of the 6 samples (S1-S6) studied in this paper is summarized in Table 1.  Cross-sectional HRTEM, shown in Fig.\ref{Figure 1} (b), reveals that, within single crystalline domains, an epitaxial interface is achieved between the GaSb and \InMnAs~layers, with grain boundaries that propagate from the \InMnAs~layer into the overgrown \CdAs~layer as indicated by the vertical dark regions and red arrows. The \CdAs~/\InMnAs~interface is of reasonable quality with some interdiffusion in a region on the order of $1-2$~nm. X-ray diffraction measurements of these films confirm growth along [001], albeit with broad rocking curves (full width at half maximum $\sim 1^{\circ}$ (see Supplementary Materials Section III). Atomic force microscopy (AFM) typically shows a root mean square surface roughness $R_q \sim 1.7$~nm and $R_q \sim 3 $~nm for \InMnAs~and \CdAs/\InMnAs, respectively (see  Supplementary Materials Section III).

\subsection*{Magnetotransport measurements}

\begin{table}[h!]
    \centering
\begin{tabular}{||c c c||} 
 \hline
Sample & Material & carrier density $ (\times 10^{13}$ cm$^{-2}$) \\ [0.5ex] 
 \hline\hline
 S1 & 25 nm \InMnAs & 17.5(p) \\ 
 \hline
 S2 & 12 nm \InMnAs & 6.79 (p) \\
 \hline
 S3 & 25 nm \CdAs & 18.6 (n) \\
 \hline
 S4 & 25 nm \CdAs~on 25 nm \InMnAs & N/A \\
 \hline
 S5 & 25 nm \CdAs~on 17 nm \InMnAs & N/A \\

 \hline
 S6 & 25 nm \CdAs~on 12 nm \InMnAs & 9.85 (n) \\
 \hline
 S5-F & Lithographically patterned Hall bar of S5 &  1.44 (n at \Vg $= 0$ V) \\
 \hline
 S6-F & Lithographically patterned Hall bar of S6 & 0.51 (n at \Vg $= 0$ V) \\
 \hline
\end{tabular}
    \caption{Samples grown and their corresponding carrier densities. All listed materials are grown on a buffer layer of 125 nm of GaSb on a 2 nm wetting layer of GaAs on a commercial GaAs (001) wafer. Electrical transport measurements for S1-S5 are carried out using a mechanically defined Hall bar. S5-F and S6-F are pieces of S5 and S6, respectively, that were fabricated into top gated Hall bars via lithography. Where shown, carrier densities and types are extracted from the slope of the high-field linear Hall effect assuming a single dominant carrier type.}
    \label{tab:1}
\end{table}

\begin{figure}
    \includegraphics[width=13cm]{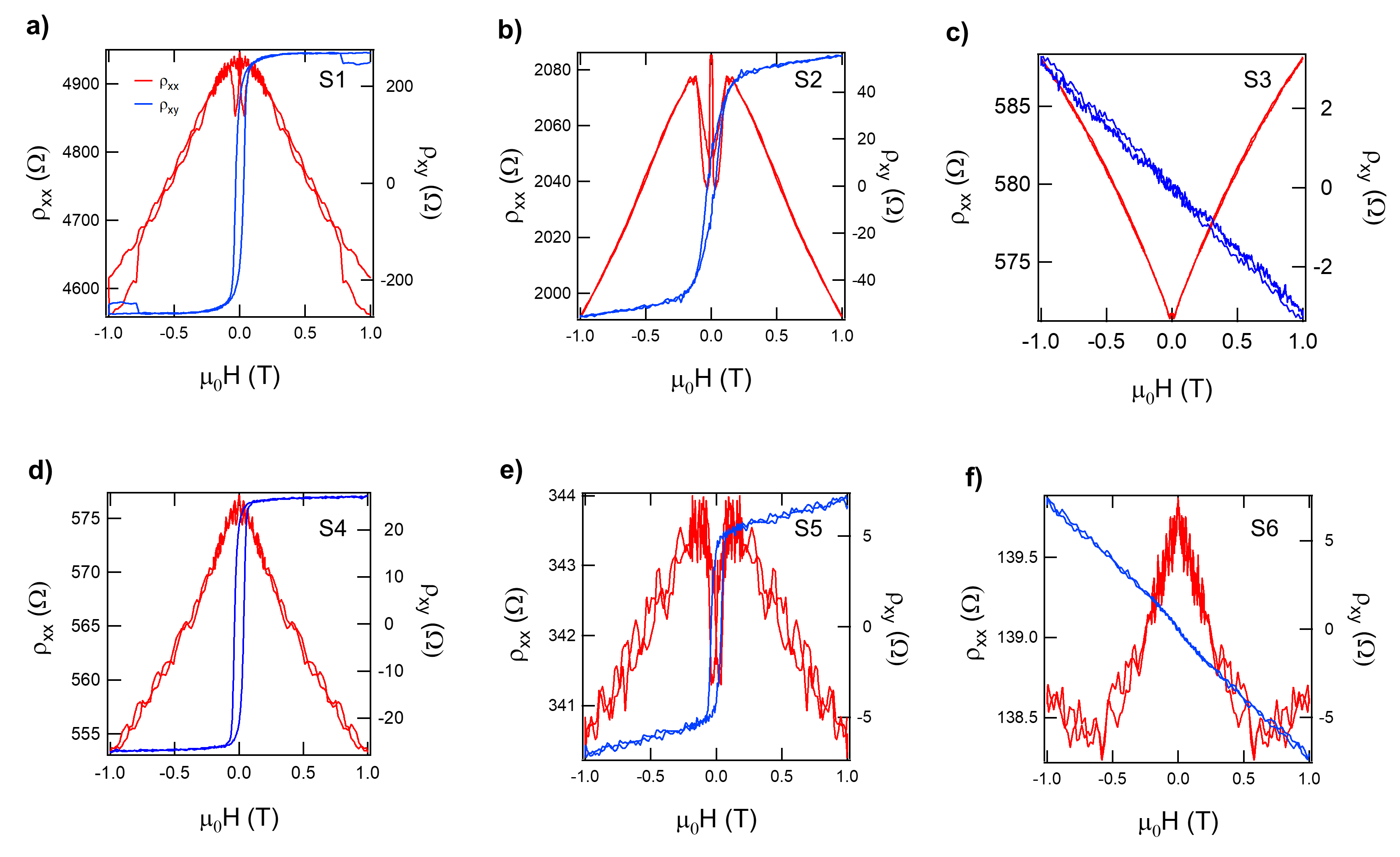}
    \caption{\textbf{Magnetotransport in control \InMnAs~and \CdAs~films and in \CdAs/\InMnAs~heterostructures:}  Upper panels show the magnetic field dependence of the 2D resistivity ( $\rho_{xx}$, left axis) and Hall resistivity ($\rho_{xy}$, right axis) in (a) S1 ($25$~nm thick \InMnAs~film),(b) S2 ($12$~nm thick \InMnAs~film), and (c) S3 ($25$~nm thick Cd$_3$As$_2$ film). The slope of $\rho_{xy}$ at high field ($\mu_0 H > 0.5 $T) indicates p-type carriers in S1 and S2, but n-type carriers in S3. Lower panels show measurements of the magnetic field dependence of $\rho_{xx}$ (left axis) and $\rho_{xy}$ (right axis) in (d) S4 ($25$~nm Cd$_3$As$_2$/25 nm \InMnAs~heterostructure, (e) S5 (25 nm Cd$_3$As$_2$/17 nm \InMnAs~heterostructure, and (f) S6 (25 nm Cd$_3$As$_2$/12 nm \InMnAs/GaSb/GaAs).}
      \label{3}
 \end{figure}

 We measure electrical magnetotransport in samples S1-S6 after mechanical patterning Hall bars of dimension $1 \times 0.5$ mm$^2$. Three of these are control epilayers of $25$~nm thick \InMnAs~(S1), $12$~nm thick \InMnAs~(S2), and $25$~nm thick \CdAs~(S3). The associated symmetrized longitudinal magnetoresistance (MR) and antisymmetrized Hall resistance in these samples are shown as a function of perpendicular magnetic field in Fig.~\ref{3} (a-c). S1 and S2 show an AHE with hysteresis in the magnetic field dependence of the Hall resistivity (\rhoxy), consistent with FM ordering with PMA in these \InMnAs~films. The amplitude of the AHE decreases with thickness of \InMnAs~(Fig. \ref{3} (b)) and disappears in samples thinner than $12$~nm. Figure \ref{3} (c) shows the symmetrized longitudinal MR and the anti-symmetrized Hall resistance of S3, a 25 nm thick \CdAs~film. We note that, at zero magnetic field, the longitudinal resistivity (\rhoxx) of the 25 nm thick \InMnAs~film (S1) is almost an order of magnitude greater than that of the 25 nm thick \CdAs~film (S3), indicating that when \InMnAs/\CdAs~heterostructures are grown, the source-drain current will be shunted by the lower resistance \CdAs~layer. Carrier densities are extracted from the linear part of \rhoxy~at high magnetic field where the AHE has saturated (See Table \ref{tab:1}). These data reveal that \InMnAs~is hole-doped ($p\sim 10^{14}$~cm$^{-2}$) while \CdAs~is electron-doped ($n\sim10^{14}$~cm$^{-2}$). The carrier density in these \CdAs~films grown on (001) GaSb/GaAs is about an order of magnitude higher than in state-of-the-art films grown under better optimized conditions \cite{kealhofer2020topological,kealhofer2021thickness}. However, as we show below, it is precisely this higher carrier density that allows us to form p-n junctions wherein the AHE can be efficiently tuned using a gate voltage. Note that Hall effect in samples S4 and S5 has comparable contributions of opposite sign from electrons and holes. Thus, even in a magnetic field range where the AHE has saturated, we cannot reliably determine the carrier density even with a two carrier fit over the magnetic field range studied here ($\mu_0 H \leq 1$T).  

Heterostructures of \CdAs/\InMnAs~are grown with $25$~nm of \CdAs~on varying thicknesses of \InMnAs~($25$~nm (S4), $17$~nm (S5) and $12$~nm (S6)). Figure \ref{3} (d-f) shows the magnetotransport in these samples. We choose to hold the thickness of \CdAs~at 25 nm to avoid the observed topological transition out of the DSM state \cite{2dticdas} and we vary the thickness of \InMnAs~to probe its influence on the heterostructure's FM state. In these heterostructures, the magnitude of \rhoxx~is on the order of that found in S3, supporting the assumption that the current preferentially flows through the \CdAs~layer. Looking at \rhoxy, we find that the AHE amplitude decreases with the thickness of \InMnAs, until it eventually disappears in S6. If we plot the Hall conductance ($\sigma_{xy}$) rather than \rhoxy, we find that the AHE amplitude in the heterostructures is more than twice as large compared to the sum of corresponding $\sigma_{xy}$ in the \CdAs~and \InMnAs~epilayers (see Supplementary Materials IV). 

\begin{figure*} 
    \includegraphics[width=12cm]{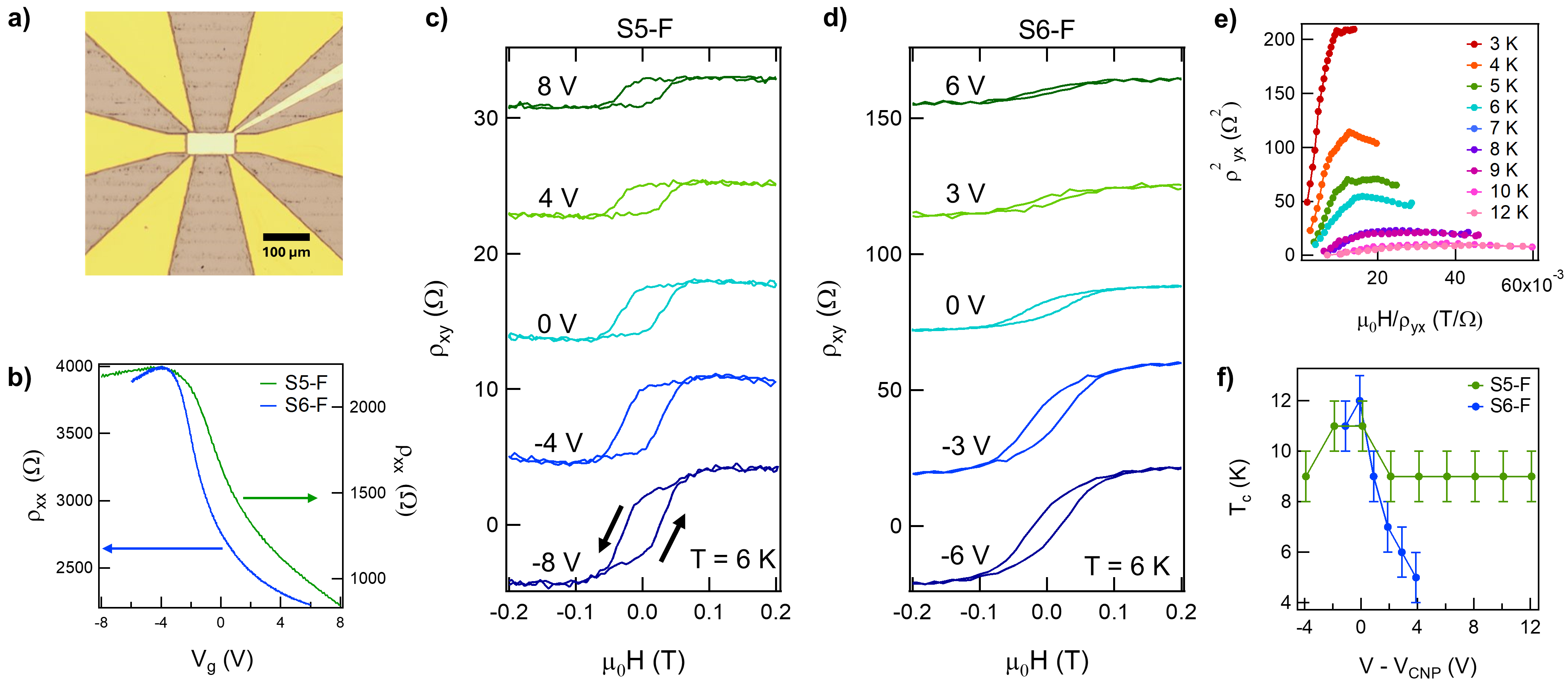}
    \caption{\textbf{Gate-voltage and temperature dependence of Hall effect:} (a) Optical micrograph of a lithographically fabricated device. The scale bar is $100$~\textmu m. (b) Longitudinal resistivity ($\rho_{xx}$) as a function of gate-voltage ($V_g$) at $T=2$~K. (c) and (d) Hall resistance ($\rho_{xy}$) in devices S5-F and S6-F, respectively, as a function of a magnetic field perpendicular to the sample plane, showing a gate-tunable AHE. The curves have been shifted vertically for clarity. (e) Arrott plot of S5-F with $V_g$ fixed at $-5$~V. (f) $T_C$ as a function of $V_g$ offset to the $V_{CNP}$. Error bars note the measurements are taken at 1 K intervals. }
    \vspace{1.5em}
    \label{4}
 \end{figure*}

To further explore the magnetic properties of these heterostructures, we measure magnetotransport in $100\times50~\mu$\textrm{m}$^2$ lithographically-patterned Hall bars (Fig. \ref{4} (a)) with an electrostatic top-gate to control the carrier density (see Methods). These devices are made from pieces of S5 and S6 and are labeled S5-F and S6-F. Figure \ref{4} (b) shows \rhoxx~as a function of the top gate voltage, $V_g$, for the two devices. Hall effect measurements in both devices show electrons as the dominant carrier type at $V_g=0$~V. The mobility of the samples is $\sim 500$~cm$^2$/Vs. Both S5-F and S6-F show maximum resistivity at $V_g \sim -4$ V and little change in resistivity at more negative gate voltage; we thus identify $V_g \sim -4$ V as the condition that positions the chemical potential at the CNP of the \InMnAs/\CdAs~heterostructure. We observe a gate-voltage-dependent AHE in both devices (Fig. \ref{3} (c, d)), where both the magnitude and width of the hysteresis loop vary with \Vg. The AHE magnitude decreases as the devices are doped with electrons ($V_g \geq 0$) in both samples and increases as the devices are doped with holes ($V_g \leq 0$). At first glance, this AHE behavior seems consistent with the hole-mediated nature of FM order in \InMnAs, superficially suggesting the magnetization in that layer as the origin of the AHE, even though the current is shunted by the lower resistance \CdAs~layer. However, as we argue later, this is not the case. Note that the AHE is almost quenched at large positive \Vg~in SF-6.

Next, we measure temperature-dependent magnetotransport to extract \Tc~in devices S5-F and S6-F using the well-established Arrott plot method \cite{arrot}. \textcolor{blue}{Figure~\ref{4} (e)} shows an example of such data at $V_g=-5$~V. The dependence of \Tc~for both samples as a function of the effective gate voltage ($V_g-V_{CNP}$) is shown in Fig.~\ref{4} (e). In the case of S6-F, we observe a sharp enhancement of \Tc~at the CNP, with \Tc~changing by almost $8$~K in the $V_g$ range that has been investigated. In S5-F, \Tc~remains constant throughout the full $V_g$ range, although we observe a small enhancement at the CNP. We attribute the difference in responses of the two devices to the higher carrier density in S5-F, resulting in screening that mitigates the gate driven changes in depletion width. 

\subsection*{Discussion}
According to the standard Zener model for FM semiconductors~\cite{Dietl_RevModPhys.86.187}, the Curie temperature (\Tc) increases monotonically with increasing hole density. This in turn affects the coercive field and the amplitude of the AHE. As mentioned above, at first sight, the observed dependence of the AHE on $V_g$ might appear to be consistent with expectations from the Zener model. As $V_g$ becomes more negative, the chemical potential decreases, significantly increasing the hole concentration and thus enhancing the \Tc, the coercivity, and the AHE amplitude (Fig. \ref{4} (c), (d)). However, as also shown in Fig. \ref{4} (e), we observe a {\it non-monotonic} change in \Tc~with possible hint of a peak at the CNP. This observation is inconsistent with the standard Zener model for FM order arising solely in the \InMnAs~layer. We also recall our earlier comment about the AHE conductance in the heterostructures noticeably exceeding the sum of the Hall conductances of the individual layers. This provides additional support for the view that the AHE is not solely due to a contribution from the InMnAs~layer. Although we currently do not have a model Hamiltonian description that can rigorously explain our observations, we propose a qualitative scenario that may account for the behavior. Prior calculations have demonstrated that Dirac electrons in a topological insulator are expected to mediate an RKKY interaction between localized magnetic spins in a vicinally interfaced material with a purely FM coupling when the chemical potential is tuned close to the Dirac point~\cite{PhysRevLett.102.156603}. We hypothesize that similar physics may play a role in the heterostructures studied here. In this case, we expect \Tc~to increase as the chemical potential is tuned towards the CNP of the DSM/FM semiconductor heterostructure. Further experiments are required to understand the role of Dirac fermion-mediated magnetism in this new family of heterostructures. We also caution that the potential role of impurity band-mediated FM order \cite{Kobayashi_PhysRevB.89.205204} needs to be considered to more fully understand our observations. 

The AHE measurements described above provide a compelling picture of the gate voltage-dependent tuning of FM order. However, other techniques are needed that more directly probe the spatial distribution of the ferromagnetism to obtain a fuller picture. As a first step towards this goal, we use polarized neutron reflectometry (PNR) measurements on an \InMnAs/\CdAs~heterostructure similar to sample S6-F, but with double the \CdAs~thickness (see Supplementary Section V). These data reveal a slight magnetization in the \CdAs~layer at the 3-$\sigma$ level and indicate that PNR has the sensitivity needed to detect a proximity effect in appropriately designed samples (e.g. with a smaller ratio of \CdAs:\InMnAs). Further iterative experiments employing techniques such as PNR on top gated samples or x-ray magnetic circular dichroism associated with element specific spin polarization are needed to fully understand the location of electrically controlled ferromagnetism in these heterostructures and the origin of the resulting AHE.


\subsection*{Conclusions}
In conclusion, we have developed a novel heterostructure that interfaces a canonical DSM, \CdAs, with a high resistivity FM semiconductor with PMA, \InMnAs, as a proof-of-concept platform for exploring gate-controllable magnetic proximity-induced Weyl fermion states in a DSM. We observe highly efficient voltage-tuning of the AHE in electrically-gated devices that indicates effective control of the hole-mediated FM order via charge transfer between the layers. The enhancement of both \Tc~and the AHE amplitude close to the CNP suggests that the gate-voltage tuning of the AHE cannot be trivially attributed to varying hole density in the \InMnAs~layer. We posit that Dirac electron-mediated ferromagnetism near the interface likely plays an important role. Two more important questions remain to be addressed in this context. First, how might one directly probe the influence of proximity-induced FM order and resulting broken TRS on the Dirac states of the \CdAs~layer? Since conventional angle resolved photoemission spectroscopy (ARPES) only probes the topmost surface, soft X-ray ARPES is likely needed to prove the conversion of the \CdAs~band structure from a DSM to WSM near a buried interface \cite{Cancellieri_PhysRevB.89.121412}. Second, can one extend our proof-of-concept demonstration of efficient gate voltage tuning of the AHE in a DSM/FM semiconductor heterostructure to technologically relevant temperatures for topological spintronics? We believe that interfacing \CdAs~with Fe-based III-V FM semiconductors \cite{Tu_10.1063/1.4948692} such as Ga$_{\textrm{1-x}}$Fe$_{\textrm{x}}$Sb offers an attractive path to this end since these FM semiconductors have \Tc~ $\sim 300$ K and are epitaxially compatible with \CdAs.

\section*{Methods}\label{sec11}

\subsection*{Band structure calculations}

Density-functional theory (DFT) calculations were conducted using the generalized gradient approximation (GGA) \cite{PhysRevLett.77.3865} of the Perdew-Burke-Ernzerhof (PBE) functional and projector-augmented wave (PAW) pseudopotentials \cite{PhysRevB.50.17953}, as implemented in VASP \cite{PhysRevB.54.11169}. Full details are given in Supplementary Materials Section I.

\subsection*{MBE growth}
The thin films and heterostructures investigated in this manuscript are grown on epiready GaAs (001) substrates in a Veeco EPI 930 MBE chamber using standard effusion cells containing As (99.999995\%), Ga (99.99999\%), Sb (99.9999\%), Mn (99.9998\%), and a high purity compound source of Cd$_3$As$_2$. Full details of the growth protocol and growth conditions as well as {\it in situ} characterization using reflection high energy electron diffraction (RHEED) are given in Supplementary Materials Section III.

\subsection*{STEM characterization}
A FEI Helios Nanolab G4 dual-beam Focused Ion Beam (FIB) system with 30 keV Ga ions was used for making cross-section samples for the STEM study. Damaged surface layers were removed using ion-milling at 2 keV and amorphous C and Pt were deposited on the surface to protect from damage on exposure to the ion beam. STEM experiments were carried out on an aberration-corrected FEI Titan G2 60–300 (S)TEM microscope, with a CEOS DCOR probe corrector, monochromator, and a super-X energy dispersive X-ray (EDX) spectrometer. A probe current of 120 pA and operation voltage of 200 keV were used for operating the microscope, and HAADF-STEM images were acquired with the probe convergence angle of 18.2 mrad with inner and outer collection angles of 55 and 200 mrad in the detector respectively. Bruker Esprit software was used to acquire and analyze EDX elemental maps.

\subsection*{X-ray diffraction}
X-ray diﬀraction patterns were collected on a 320.00 mm radius Malvern Panalytical X’Pert3 MRD four circle X-ray diﬀractometer equipped with a line source [Cu K-$\alpha$ 1-2 (1.5405980/ 1.5444260 $\AA$)] X-ray tube at 45.0 kV and 40.0 mA. The incident beam path included a 2xGe(220) asymmetric hybrid monochromator with a 1/4° divergence slit. A PIXcel3D 1x1 detector operating in receiving slit mode was used with an active length of 0.5 mm. PHD lower and upper levels were set at 4.02 and 11.27 keV respectively. $\Phi$ scans were performed on the same instrument using the PIXcel3D 1x1 detector operating in open detector mode. Analysis was carried out using Jade®software (version 9.1) from Materials Data Inc. (MDI) and the International Centre for Diﬀraction Data (ICDD) PDF5 database.

\subsection*{Magnetotransport measurements}
Magnetotransport  measurements are performed on Hall bar devices using a Quantum Design Dynacool Physical Properties Measurement System. The devices are either mechanically scribed to $1 \times 0.5$ mm$^2$ Hall bars or lithographically patterned into Hall bars of dimension $100\times50~\mu$\textit{m}$^2$ with an electrostatic top-gate (30 nm Au on 10 nm Cr) and a dielectric spacer (30 nm $\mathrm{Al_{2}O_{3}}$).

\subsection*{Polarized neutron reflectometry}

 Polarized neutron reflectivity (PNR) measurements of a sample with nominal structure Au (30 nm)/Cr (10 nm)/$\mathrm{Al_{2}O_{3}}$ (28 nm)/\CdAs~ (56 nm)/\InMnAs~ (12 nm)/GaSb (180 nm)/GaAs were performed at 7 K in an in-plane 0.7 T magnetic field on the MAGREF reflectometer at the Spallation Neutron Source at Oak Ridge National Laboratory (ORNL). Full details of the PNR measurements and analysis are given in Supplementary Materials Section IV

\section*{Data Availability}\label{sec12}
All data for the figures and other Supplementary information that support this work are available upon reasonable request to the corresponding author. These data will also be publicly available at (link to be added). 

\section*{Ethics Declarations}
The authors declare no competing interests

\section*{Acknowledgments}
This work was primarily supported by grant number NSF DMR-2407130 (ES, SI, NS). We also acknowledge partial support for this project, including the use of low-temperature transport facilities (DOI: 10.60551/rxfx-9h58), from the Pennsylvania State University Materials Research Science and Engineering Center [grant number NSF DMR-2011839] (SI, ES, NS). ES acknowledges support from the National Science Foundation Graduate Research Fellowship Program under Grant No. DGE1255832. AC and CXL acknowledge the support of the NSF grant award (DMR-2241327). BN and YW acknowledge startup funds from the University of North Texas, and computational resources from the Texas Advanced Computing Center. Part of the modeling was supported by computational resources from a user project at the Center for Nanophase Materials Sciences (CNMS), a US Department of Energy, Office of Science User Facility at Oak Ridge National Laboratory. DISCLAIMER
We identify certain commercial equipment, instruments, and materials in this article to specify adequately the experimental procedures. In no case does such identification imply recommendation or endorsement by the National Institute of Standards and Technology nor does it imply that the materials or equipment identified are necessarily the best available for the purpose. This project was initiated in part using support from the Institute for Quantum Matter under DOE EFRC Grant No. DESC0019331 (JC, TM, NS). SG and KAM acknowledge support by NSF Grant No. DMR-2309431. STEM portion of this work was carried out in the Characterization Facility, University of Minnesota, which receives partial support from the NSF through the MRSEC (DMR-2011401).

\section*{Author Information}

\subsection*{Authors and Affiliations}

\subsection*{Author Contributions}
N.S. conceived the project. E.S. grew the samples by MBE, with technical assistance from K.A.N. E.S. patterned the electrical transport devices and performed magnetotransport measurements with guidance from S.I. W.J.Y.P carried out x-ray diffraction characterization. S.G. conducted TEM characterization and analyzed the TEM data under the supervision of K.A.M. B.N. and A.C. carried out theoretical calculations under the supervision of Y.W. and C.X.L. A.G., J.B., C.J. and T.C. carried out PNR measurements and the related analysis. J.C. and T.M. synthesized the high purity \CdAs~compound source material. E.S., S.I., and N.S. wrote the manuscript with substantial contributions from all authors.
\subsection*{Corresponding Authors}

\bibliography{ES_refs}

\end{document}